\documentclass[floatfix, twocolumn, preprintnumbers, amsmath, amssymb, prl, superscriptaddress]{revtex4-2} 
\usepackage{url}
\usepackage{graphicx}
\usepackage{dcolumn}
\usepackage{xspace}
\usepackage{bm}
\usepackage{soul}
\usepackage{ulem}
\usepackage{xcolor}
\usepackage{color}
\usepackage{natbib}
\usepackage[mathlines]{lineno}
\usepackage{ulem}

\newcommand{\pt}{\mbox{$p_{\rm{T}}$}\xspace}
\newcommand{\snn}{\mbox{$\sqrt{s_{\rm NN}}$}\xspace}
\newcommand{\hyt}{\mbox{${}^3_\Lambda \rm{H}$}\xspace}

\newcommand{\sNN}{\sqrt{s_{\rm NN}}}
\newcommand{\he}{{}^{3}{\rm He}}
\newcommand{\mpt}{\langle p_{\rm T} \rangle}

\begin{document}
\title{
Collision Energy Dependence of Hypertriton
Production in Au+Au Collisions at RHIC
}
\author{The STAR Collaboration}
\date{\today}

\begin{abstract}
The STAR Collaboration reports measurements of the collision energy dependence of hypertriton (${}^{3}_{\Lambda}$H) transverse momentum spectra and $\pt$-integrated yields at mid-rapidity ($|y|<$0.5) in Au+Au collisions at 11 collision energies between 3.2 and 27\,GeV. The measured ${}^{3}_{\Lambda}$H yields and ${}^{3}_{\Lambda}$H/$\Lambda$ yields ratio in central collisions increase strongly with decreasing collision energy, and are a factor of $\sim$2 lower than thermal model predictions at this energy range. The mean $\pt$ of ${}^{3}_{\Lambda}$H is lower than the Blast-Wave expectation using the freeze-out parameters from light hadrons. Furthermore, the observed double ratio $({}^{3}_{\Lambda}{\rm{H}}/\Lambda)/(t/p)$ maintains a constant value of $\sim$0.4 across the measured energy range. Within the coalescence framework, this ratio directly reflects the significantly suppressed formation probability of the weakly-bound hypertriton relative to the triton, which results from the weaker hyperon-nucleon interaction compared with the nucleon-nucleon interaction.
\end{abstract}

\maketitle

Relativistic heavy-ion collisions offer a unique laboratory to study the strong interaction under extreme temperature and density environments as well as emergent properties governed by Quantum Chromodynamics (QCD). 
At high baryon densities, two-body and three-body interactions between nucleons ($N$) and hyperons ($Y$) contribute significantly to the Equation-of-State (EoS) of QCD matter~\cite{Gal:2016boi,Tolos:2020aln,Burgio:2021vgk} and can offer significant insights into understanding the structure of compact astrophysical objects~\cite{Lonardoni:2014bwa,Gerstung:2020ktv,Burgio:2021vgk}. Hypernuclei, bound states composed of nucleons and hyperons, are excellent probes to study the $Y$-$N$ interaction.

Due to the higher baryon densities at lower collision energies, theoretical calculations from thermal models and coalescence models predict a significant enhancement of light nuclei ($A<4$) and hypernuclei yields at low collision energies  (i.e., $\snn<20$~GeV) compared to those at the top RHIC energy ($\snn=200$~GeV)~\cite{Andronic:2010qu,Steinheimer:2012tb,Glassel:2021rod,Reichert:2022mek}.
In thermal model calculations, yields of all particles are computed assuming thermal equilibrium at chemical freeze-out, at which point all hadron yields are frozen \cite{Braun-Munzinger:2003pwq,Andronic:2017pug}. It is argued that this picture can also hold for loosely bound hypernuclei because the relative yields of all nuclei are determined by the entropy per baryon, also fixed at chemical freeze-out and conserved thereafter~\cite{Andronic:2010qu,Andronic:2017pug}. In contrast, coalescence models assume that hypernuclei are formed through the binding of nucleons and hyperons close to the kinetic freeze-out of hadrons, the stage at which all scatterings cease and the momentum distributions of particles are fixed \cite{Schnedermann:1993ws,STAR:2017sal}. Precise experimental data on hypernuclei yields are crucial for distinguishing between these models, and for providing insights into the role of the $Y$-$N$ interaction in hypernuclei formation.

Hypertriton ($\hyt$), the lightest bound state of hypernuclei, consists of a proton, a neutron, and a $\Lambda$ hyperon. Numerous measurements have been made on $\hyt$ intrinsic properties from heavy-ion collision experiments \cite{Chen:2023mel}, e.g. $\Lambda$ separation energy ($B_{\Lambda}$) \cite{STAR:2019wjm,ALICE:2022sco}, lifetime \cite{STAR:2021orx, ALICE:2019vlx,ALICE:2022sco} and decay branching ratio \cite{Ji:2023fqp, STAR:2017gxa}. In particular, its $\Lambda$ separation energy has been measured to be between $0.07$ and $0.41$\,MeV \cite{STAR:2021orx,ALICE:2022sco}, $\sim$10 times smaller than the binding energy of triton (${}^{3}\mathrm{H}$, denoted as $t$). In comparison, there are fewer measurements on the yields of hypernuclei, and they are limited to either very low ($\snn\leq5$\,GeV) \cite{STAR:2021orx,RAPPOLD2015129} or very high collision energies ($\snn\geq200$\,GeV) \cite{STAR:2010gyg, STAR:2023fbc, ALICE:2015oer, ALICE:2021puh}, often with considerable uncertainties.

Recently, the STAR experiment observed a significant directed flow~\cite{STAR:2014clz} of hypernuclei ($\hyt$ and ${}^{4}_{\Lambda}{\rm H}$) at $\snn=3$\,GeV~\cite{STAR:2022fnj}. The slopes of the directed flow against rapidity for hypernuclei at mid-rapidity scale with mass, similar to that observed for nuclei. This observation is qualitatively consistent with the formation of hypernuclei and nuclei via the coalescence mechanism. 
However, the microscopic picture of the coalescence process, especially its dependence on the (hyper-)nuclei properties (binding energy, radii etc.) as well as the high-density environment created in these collisions, remains elusive.
A comprehensive measurement of $\hyt$ yields over a broad range of collision energies would be essential for understanding the $\hyt$ production mechanisms in heavy-ion collisions and for providing a crucial benchmark for measuring heavier hypernuclei in future experiments~\cite{Herrmann:2022jkv, Syresin:2019vzo, Aoki:2021cqa, Zhou:2022pxl}.

In this letter, we report measurements of hypertriton production in Au+Au collisions at $\snn=$ 3.2, 3.5, 3.9, 4.5, 5.2, 7.7, 11.5, 14.6, 17.3, 19.6, and 27~GeV utilizing the high-statistics data collected during the second phase of the STAR Beam Energy Scan program (BES-II) at RHIC \cite{Yang:2019bjr}. The data from $\snn=3.2$ to $5.2$~GeV were taken using the fixed-target (FXT) configuration~\cite{STAR:2021yiu}, while those from $7.7$ to $27$~GeV were taken using the collider (COL) configuration. For the FXT configuration, a gold target was placed inside the beam pipe at the western edge of the Time Projection Chamber (TPC)~\cite{Anderson:2003ur, WANG201790}, 201 cm away from the center of the STAR detector~\cite{STAR:2021yiu}.

For the COL configuration, the minimum-bias (MB) trigger~\cite{Judd:2018zbg} requires the coincidence of signals from the east and west Zero Degree Calorimeters (ZDCs) \cite{Xu:2016alq}, Vertex Position Detectors (VPDs)~\cite{Llope:2014nva}, or Beam-Beam Counters (BBCs)~\cite{Whitten:2008zz}. Presence of signal hits in the Time-of-Flight (TOF)~\cite{BONNER2003181} is also required when the BBC is used as the triggering detector. For the FXT configuration, the MB trigger is provided by signals in the east BBC and at least five hits in the TOF detector. For the FXT analyses, the reconstructed primary vertex of each event is required to be within $198<V_{z}<202$ cm along the beam direction and within a radius of 1.5 cm from the nominal primary vertex in the radial plane. For the COL analyses, the distance between the primary vertex and the nominal center of the STAR detector is required to be within $\pm70$ cm along the beam direction and $2$ cm in the radial direction.  
In total, 214, 117, 116, 129, 89, 58, 173, 200, 188, 421, and 521 million qualified events at $\snn=$ 3.2, 3.5, 3.9, 4.5, 5.2, 7.7, 11.5, 14.6, 17.3, 19.6, and 27~GeV are used.
The centrality of each collision is determined by comparing the measured charged particle multiplicity in a certain kinematic region ($|\eta|<0.5$ for COL configuration and $-2<\eta<0$ for FXT configuration) to a Glauber model Monte Carlo simulation \cite{Miller:2007ri, STAR:2009sxc}. 

\begin{figure}[htb]
\includegraphics[width=0.9\columnwidth]{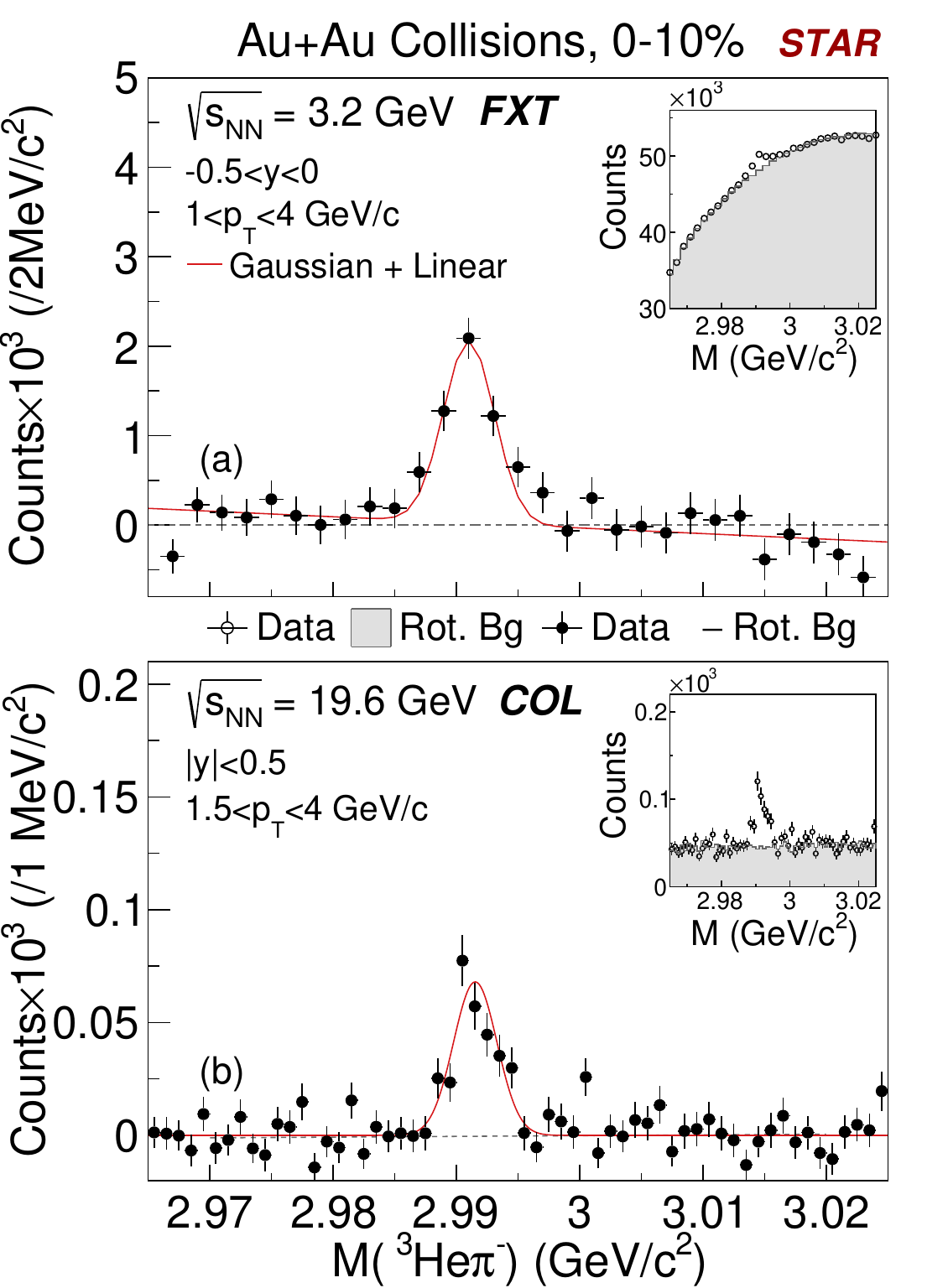}
\caption{\label{fig:signal}
Combinatorial background subtracted invariant mass distributions of $\rm {}^{3}_{\Lambda}H\rightarrow {}^{3}He\pi^{-}$ candidates in 0-10\% central Au+Au collisions at $\sqrt{s_{\rm NN}}$ =  3.2 (top) and 19.6\,GeV (bottom), with un-subtracted distributions inset. They are fitted with a Gaussian plus a linear function, and the fit results are shown as red curves. Invariant mass distributions before combinatorial background subtraction (open circles) and combinatorial background distributions themselves as estimated by rotating $\rm {}^{3}He$ tracks (gray filled area) are shown in the insets.
}
\label{fig:signal}
\end{figure}

The TPC is the major detector subsystem for charged-particle track reconstruction. To ensure good track quality for the analyses, tracks are required to leave at least 15 hit points out of a maximum of 72 in the TPC. Particle species are identified based on the correlation between their average ionization energy loss per unit length ($\langle dE/dx \rangle$) in the TPC and $p/q$, where $p$ and $q$ are the momentum and charge of a particle, respectively. Tracks are required to have more than 5 TPC hit points used for the $\langle dE/dx \rangle$ calculation.

The $\hyt$ candidates are reconstructed via the $\hyt\rightarrow\he\pi^{-}$ decay channel using the KFParticle package \cite{Zyzak:2016exl,Ju:2023xvg}. Selection criteria on various decay topology variables provided by KFParticle are applied to optimize the signal significance. The combinatorial background is estimated via a rotation technique, where the $\he$ track in each event is rotated by a random angle between $10^{\circ}$ and $350^{\circ}$. Resulting invariant mass distributions with rotated $\he$ tracks are normalized to match the data distributions within the mass regions $(3.008, 3.018)$ for FXT analyses and $(2.970, 2.983)\cup(3.000, 3.020)$ GeV/$c^{2}$ for COL analyses. After subtracting the combinatorial background, the invariant mass distributions of the $\hyt$ candidates are fitted with a Gaussian function plus a linear function, which describes the signal and residual background, respectively. Finally, the integral of the Gaussian function is taken as the raw signal count ($N^{\rm raw}$).

Figure~\ref{fig:signal} shows the invariant mass distributions of $\he\pi^{-}$ pairs in 0-10\% central Au+Au collisions at $\snn=$ 3.2 and 19.6~GeV. The track momentum resolution in the COL configuration is better than that in the FXT configuration, leading to a narrower $\hyt$ invariant mass peak for COL energies.

\begin{figure}[htb]
\includegraphics[width=0.85\linewidth]{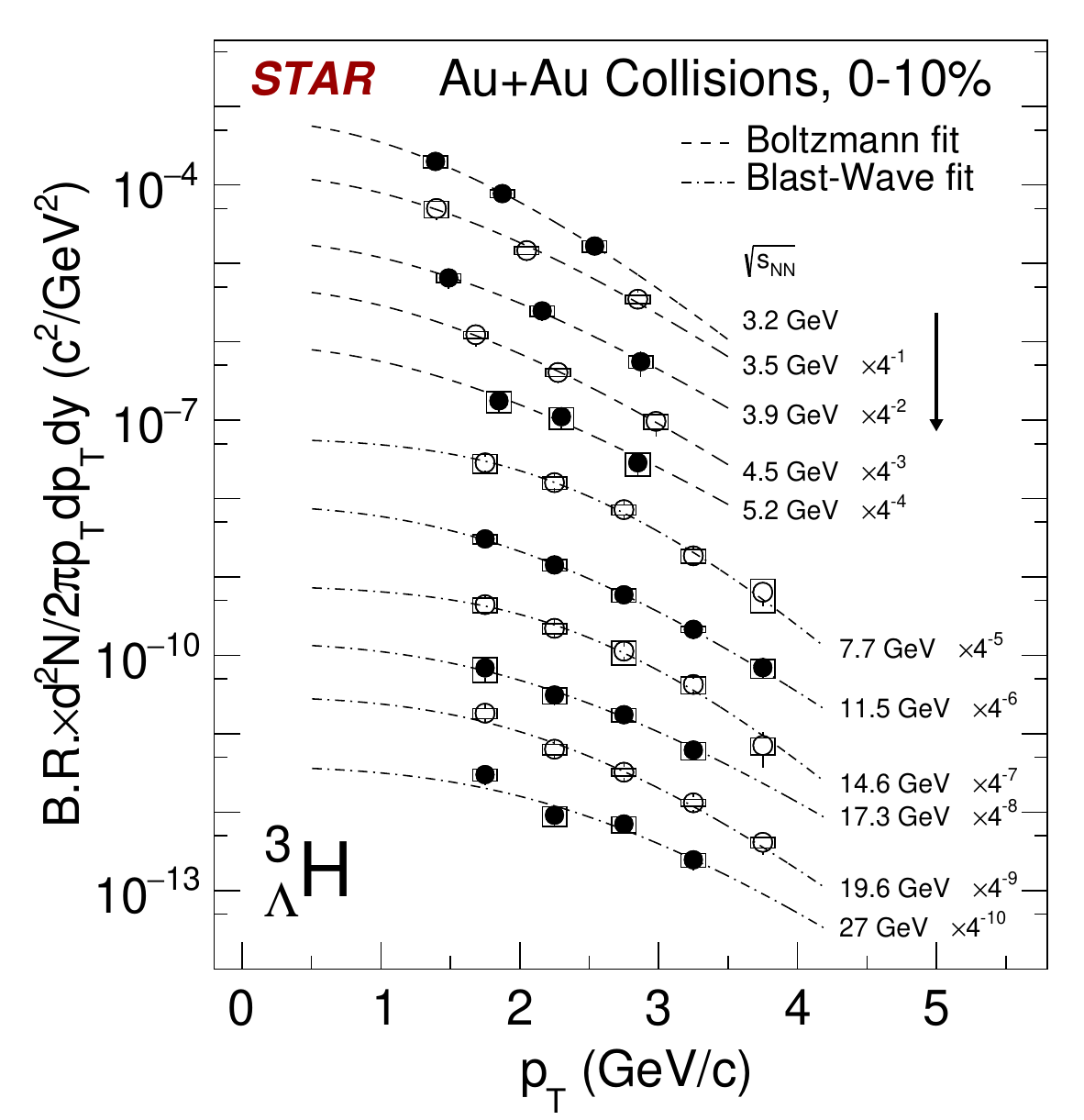}\caption{\label{fig:ptspectra} Transverse momentum spectra of $\rm {}^{3}_{\Lambda}H$ from $\sNN=$~3.2 to 27 GeV in 0-10\% central Au+Au collisions at mid-rapidity. The spectra are measured in the rapidity ranges of $-0.5<y<0$ for $\sNN = 3.2$–$4.5$ GeV, $-0.8<y<-0.3$ for $\sNN = 5.2$ GeV, and $-0.5<y<0.5$ for $\sNN = 7.7$–$27$ GeV. The dashed and dot-dashed lines are the Boltzmann and Blast-Wave fits to the data, respectively. The boxes indicate systematic uncertainties, while the vertical bars represent statistical uncertainties.}
\label{fig:ptspectra}
\end{figure}

Throughout this paper, rapidities are always given in the center-of-mass frame unless otherwise specified. The $\hyt$ yields are measured within the rapidity range of $|y|<0.5$ for the COL energies, and $-0.5<y<0$ $(-0.8<y<-0.3)$ for the FXT energies $\sqrt{s_{\rm{NN}}}=3.2-4.5$ GeV ($\sqrt{s_{\rm{NN}}}=5.2$ GeV). The rapidity coverages in the COL and FXT modes (except for $\sqrt{s_{\rm{NN}}}=5.2$ GeV) correspond to the same midrapidity region due to the symmetry of the collider system.
The $\hyt$ invariant yield is computed as:
\begin{equation}
    B.R.\times\frac{d^2 N}{2\pi\pt d\pt dy} = \frac{1}{2\pi \cdot N^{\rm evt}\cdot A \varepsilon^{\rm reco}} \cdot \frac{N^{\rm raw}}{\pt\Delta \pt \Delta y },
\label{eq1}
\end{equation}
Here, $N^{\rm evt}$ is the number of events. $\Delta \pt$ and $\Delta y$ are the widths of the transverse momentum ($\pt$) and rapidity intervals. $B.R.$ is the branching ratio of $\hyt\rightarrow\he\pi^{-}$, which is estimated to be $23\pm3\%$ (see End Matter). The left side of the Eq.~\eqref{eq1} is expressed as $B.R.$ times yields to avoid the need to include the uncertainty on $B.R.$ in the reported yields. $A\varepsilon^{\rm{reco}}$ is the product of $\hyt$ acceptance and reconstruction efficiency, which includes contributions from detector acceptance, the track reconstruction efficiency, the particle identification efficiency, and the efficiency of the selection cuts on the topological variables. They are all evaluated using the standard STAR embedding technique \cite{STAR:2017sal,STAR:2019bjj}, where
$\hyt\rightarrow{\rm {}^{3}He}\pi^{-}$ decays are simulated and the decay products are transported through a GEANT3 \cite{Brun:1987ma} simulation of the STAR detector. The simulated signals are then mixed with the digital signals from real data and reconstructed using the same algorithm and selection cuts as in the data analyses. 
Figure \ref{fig:ptspectra} shows fully-corrected $\hyt$ $\pt$ spectra times the decay branching ratio at mid-rapidity in 0-10\% Au+Au collisions from $\snn=$ 3.2 to 27 GeV. 
The $\pt-$integrated $\hyt$ yields ($dN/dy$) are obtained by extrapolating the measured $\pt$ spectra to the unmeasured regions using the Boltzmann function~\cite{STAR:2017sal} for FXT configuration:
\begin{equation}
 C\cdot m_{\rm T}\exp(-\frac{m_{\rm T}}{T}),
 \label{eq2}
\end{equation}
 where $m_{\rm T}=\sqrt{m_{0}^{2}+p_{\rm T}^{2}}$ and $m_{0}$ is the mass of $\hyt$. For COL, the Blast-Wave function is used (see its form in End Matter)~\cite{Schnedermann:1993ws,STAR:2017sal}. These fit functions are also used to extract the mean \pt $(\langle \pt \rangle)$.

Sources of systematic uncertainties on the $\pt$ spectra include those from raw yield extraction, and the acceptance-and-efficiency correction. 
The uncertainties in the raw yield extraction are evaluated by varying the fitting range of the background-subtracted invariant mass distributions and by applying a bin-counting method~\cite{STAR:2021hyx} as an alternative yield extraction method. For the systematic uncertainty in the acceptance and efficiency correction, the following sources are considered: (1) the track reconstruction efficiency; (2) the lifetime of the $\hyt$ in the simulation; and (3) mismatches between data and simulation in topological variable distributions. They are estimated by varying the requirement on the minimum number of TPC hits for a track, varying the lifetime of the simulated \hyt by $\pm1\sigma$ around the experimental average ($\tau(\hyt)=228\pm12$~ps is used in the analyses~\cite{Eckert:2022dyz}), and varying the criteria for topological variables used in candidate selection. The Barlow method is used to remove the eﬀects of statistical fluctuations
in the systematic uncertainty estimation \cite{Barlow:2002yb}. For $dN/dy$ and $\langle \pt \rangle$, the uncertainty due to the extrapolation into the unmeasured region is considered as well. Four functional forms \cite{STAR:2021orx,STAR:2021hyx}, including Blast-Wave function \cite{STAR:2017sal}, the Boltzmann function Eq. \eqref{eq2}~\cite{STAR:2017sal}, $C\cdot \exp(-\frac{m_T}{T})$ \cite{STAR:2017sal}, and $C\cdot\exp(-\frac{\pt^{2}}{\mu})$ \cite{STAR:2021orx,STAR:2021hyx} , are tried for extrapolation; and half of the maximum differences in $dN/dy$ and $\langle \pt \rangle$ obtained using these functions are taken as the systematic uncertainties. Finally, different sources of systematic uncertainties are assumed to be uncorrelated and added in quadrature. The global systematic uncertainty from the $\hyt$ decay branching ratio is quoted separately, and is not included in the $dN/dy$ systematic uncertainties.

\begin{figure}[htb]
\includegraphics[width=0.9\columnwidth]{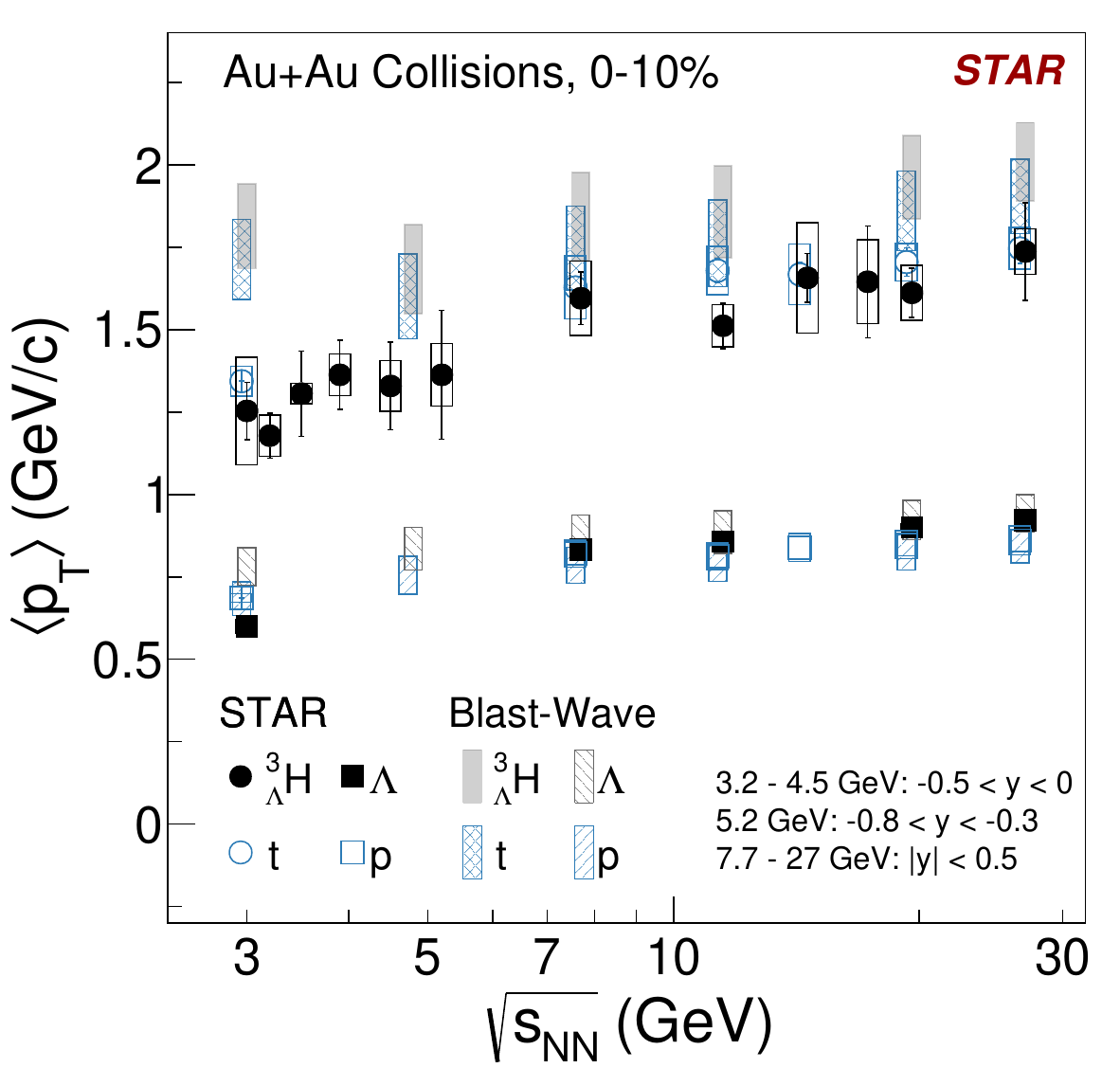}
\caption{\label{fig:meanpt} $\langle p_{\rm T} \rangle$ of $\rm {}^{3}_{\Lambda}H$ in 0-10\% central Au+Au collisions as a function of collision energy. $\langle p_{\rm T} \rangle$ of $\Lambda$ (solid squares) \cite{STAR:2019bjj, STAR:2024znc}, $t$ (open circles) \cite{STAR:2022hbp,STAR:2023uxk}, and $p$ (open squares) \cite{STAR:2017sal,STAR:2023uxk} measured by STAR are also shown for comparison. Data points for $t$ and $p$ are shifted horizontally for clarity. The boxes represent systematic uncertainties and the vertical bars represent statistical uncertainties. The expectations of the $\langle p_{\rm T} \rangle$ for $\rm {}^{3}_{\Lambda}H$ (gray non-hatched), $\Lambda$ (gray hatched), $t$ (blue non-hatched), and $p$ (blue hatched), assuming a Blast-Wave parametrization based on the measured kinetic freeze-out parameters from light hadrons, are shown as colored bands, where the vertical extents of the bands represent their total uncertainties.} 
\label{fig:meanpt}
\end{figure}

\begin{figure}[htb]
\includegraphics[width=0.8\columnwidth]{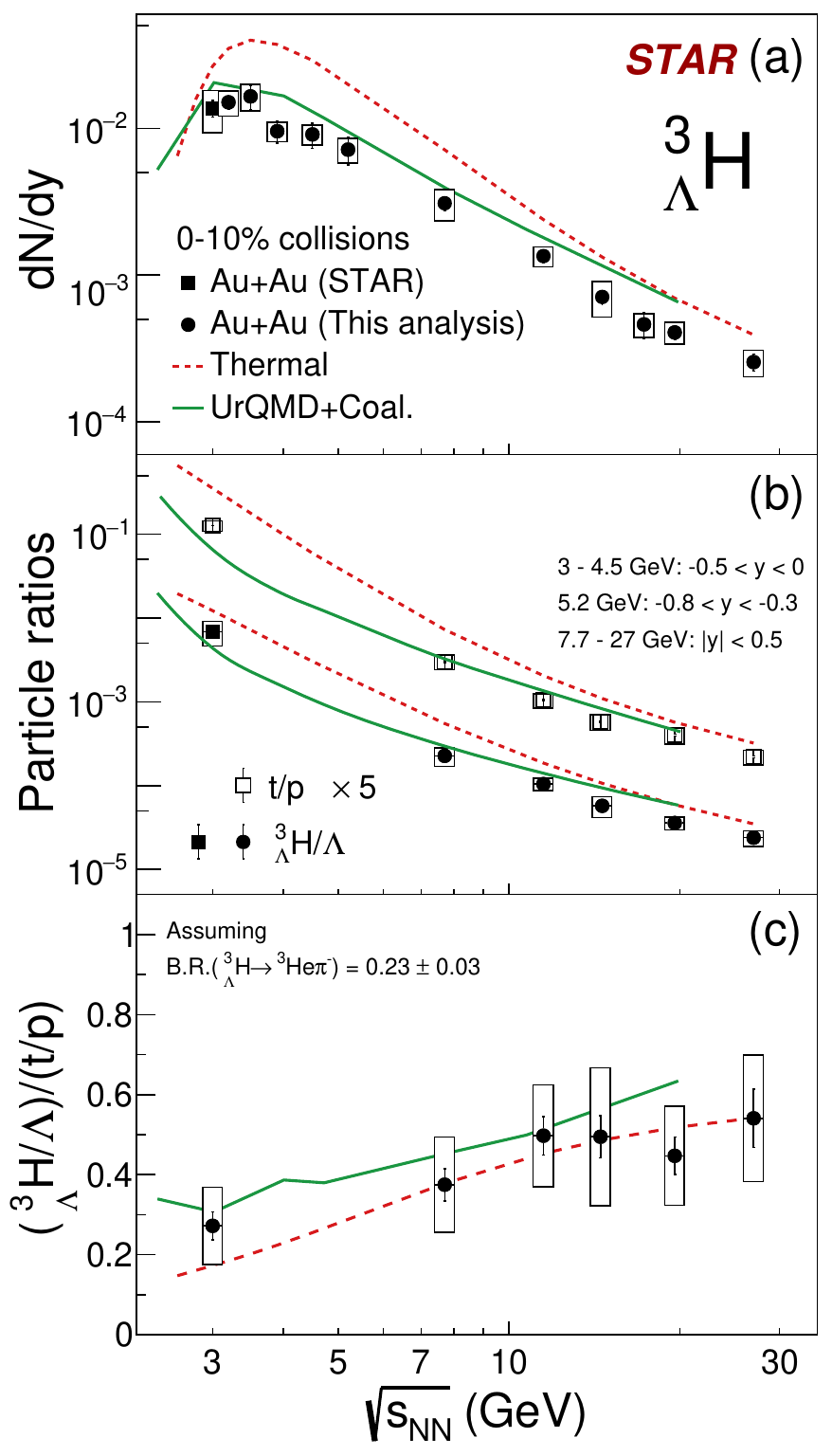}
\caption{\label{fig:dndy} The collision energy dependence of ${}^{3}_{\Lambda}\rm{H}$ yields (top panel) \cite{STAR:2021orx}, ${}^{3}_{\Lambda}\rm{H}/\Lambda$ and $t/p$ \cite{STAR:2023uxk,STAR:2022hbp} yields ratio (middle panel), as well as $\hyt/\Lambda$ to $t/p$ ratio (bottom panel). The dashed curves are from thermal model calculations~\cite{Vovchenko:2015idt}, while dotted curves are based on the UrQMD model with the coalescence of hyperons and nucleons as an afterburner \cite{Reichert:2022mek}. Boxes around data points indicate systematic uncertainties, and the vertical bars represent statistical uncertainties. The 13\% uncertainty in the ${}^{3}_{\Lambda}\rm{H}$ branching ratio is not shown.}
\label{fig:dndy} 
\end{figure} 

Figure \ref{fig:meanpt} shows the $\langle\pt\rangle$ of $\hyt$ as a function of $\snn$ in 0-10\% Au+Au collisions. Data exhibit a mild increasing trend with increasing collision energy. $\langle \pt \rangle$ of $\Lambda$~\cite{STAR:2019bjj, 
STAR:2024znc}, $t$~\cite{STAR:2022hbp}, and $p$~\cite{STAR:2017sal} are also shown for comparison. We observe that $\langle \pt \rangle$ of \hyt is consistent with that of $t$ from $3$ to $27$ GeV, but substantially larger than those of the lighter $p$ and $\Lambda$, indicating that $
\langle \pt \rangle$ is largely driven by particle mass.

To further investigate their production dynamics, we calculate the $\langle \pt \rangle$ for $p,~\Lambda,~t,~\hyt$, assuming that they follow the same hydrodynamical expansion as light-flavor hadrons, i.e., $\pi$, $K$, $p$. Specifically, we assume that the $\pt$ spectra of (hyper)nuclei are described by Blast-Wave functions with a common freeze-out temperature and flow velocity as for $\pi$, $K$, $p$~\cite{STAR:2017sal, STAR:2023uxk, E802:1999hit}, which are compiled in Tab. 2 of End Matter. The Blast-Wave parameterized mean $p_{\rm T}$, $\mpt^{\rm BW}$, are shown as colored bands in Fig. \ref{fig:meanpt}. The vertical extents of the bands represent the uncertainties arising from those in the freeze-out parameters. The proton $\langle \pt \rangle$ is consistent with $\mpt^{\rm BW}$ by construction. For \hyt, all data tend to lie below the Blast-Wave expectation, especially at $\sqrt{s_{\rm{NN}}}=3$ GeV, where the measured $\langle \pt \rangle$ is smaller than $\mpt^{\rm BW}$ by approximately $0.5$ GeV$/c$. This indicates that $\hyt$ and $t$ do not follow the same collective radial expansion as the light hadrons, particularly at energies around 3 GeV.

Figure \ref{fig:dndy} (a) shows the collision energy dependence of $\hyt$ $dN/dy$ at mid-rapidity in 0-10\% Au+Au collisions. The $\hyt$ yields increase sharply from $\snn=$ 27 GeV down to 4.5 GeV and attains a maximum at $\snn$ around 3--4~GeV. There are two competing effects as the collision energy decreases: the baryon density rises, which implies an enhanced number of baryons at mid-rapidity, and the strange hadron yields decrease, especially when the collision energy is close to the strange hadron production thresholds. Panel (b) of Fig.~\ref{fig:dndy} shows the yields ratio $\hyt/\Lambda$ as a function of collision energy. It decreases steeply from $\snn = 3$ to 27 GeV, similar to the trend observed for the $t/p$ yields ratio, which is also shown in the panel. Since both $\hyt$ and $\Lambda$ contain one strange quark, the strangeness-related suppression largely cancels in the $\hyt/\Lambda$ ratio. As a result, the ratio is primarily sensitive to the baryon density, leading to a collision-energy dependence similar to that of $t/p$.

The results are compared to thermal model calculations including unstable nuclear states (see End Matter for detailed settings). The $\hyt$ yields, $\hyt/\Lambda$ yields ratio, and $t/p$ yields ratio are systematically overestimated by the thermal model calculations by approximately a factor of two in the reported energy range. This is different from deuterons, whose yields can be well described by the thermal model~\cite{STAR:2019sjh}. This implies, within a thermal-model framework, that $\hyt$ and $t$ do not share the same freeze-out surface as that of light hadrons and deuterons. Similar observations for $t$ and ${}^{3}\rm{He}$ were reported by ALICE in Pb+Pb collisions at 5.02 TeV~\cite{ALICE:2022veq}. 
A suppression of the $A=3$ nuclei yields with respect to the thermal model predictions may arise from hadronic re-scatterings in the late hadronic stage, as suggested in Ref.~\cite{Sun:2022xjr}. It is found that, after including the dissociation and regeneration reactions ($\pi NN \leftrightarrow \pi d$ and 
$\pi NNN \leftrightarrow \pi\, {}^{3}\mathrm{H}$(${}^{3}$He)) during the hadronic stage within a kinetic approach, the $A=3$ nuclei ($t$ and ${}^{3}$He) yields are reduced 
by about a factor of 1.8 compared to the thermal model predictions, while the 
$d$ yields remains largely unaffected.

We also compare our results with coalescence calculations~\cite{Reichert:2022mek} (solid curves in Fig.~\ref{fig:dndy}), which provide a better description of the data than the thermal model. These calculations use the UrQMD model~\cite{Bass:1998ca, Bleicher:1999xi} to produce nucleons and hyperons, which then form nuclei and hypernuclei via coalescence when their relative momentum ($\Delta p$) and distance ($\Delta R$) fall below defined thresholds. For ${}^{3}_{\Lambda}$H, $\Delta p < 0.135$ GeV$/c$ and $\Delta R < 9.5$ fm, while for triton, $\Delta p < 0.33$ GeV$/c$ and $\Delta R <4.3$ fm. These thresholds approximate the uncertainty principle ($\Delta p \Delta R \approx \hbar$). The ${}^{3}_{\Lambda}$H’s looser spatial constraint and stricter momentum requirement than triton reflect its wider wavefunction, stemming from the weaker hyperon-nucleon ($Y$–$N$) binding compared to the stronger nucleon-nucleon ($N$–$N$) interaction in the triton. As demonstrated in Ref.~\cite{Reichert:2022mek}, these constraints, arising from the interplay between its extended wavefunction and the momentum-space correlations of nucleons and hyperons, inherently reduce the hypertriton coalescence probability relative to the triton. The coalescence calculations qualitatively reproduce the trend observed in the data, although they tend to overestimate the measurements at $\snn >10$ GeV.

Figure~\ref{fig:dndy} (c) shows the double ratio $(\hyt/\Lambda)/(t/p)$, which is proposed to be sensitive to the coalescence probability ratio for $\hyt$ and $t$~\cite{Dong:2018cye, Sun:2018mqq}, as a function of collision energy. The observed ratio is consistent with a constant value of approximately 0.4 over the measured energy range, and is well described by both thermal and coalescence models. 
Within the coalescence framework, the deviation from unity can be interpreted as a reduced coalescence probability for the \hyt, which will provide constraints on $Y$-$N$ interactions when combined with an accurate modelling of the kinematic distributions for nucleons and hyperons~\cite{Bellini:2020cbj, Mahlein:2025bla}.

In summary, STAR reports the energy dependence of hypertriton (${}^3_\Lambda \rm{H}$) production yields ($dN/dy$) at mid-rapidity in 0-10\% Au+Au collisions across the collision energy range of 3.2--27\,GeV. The measured ${}^3_\Lambda \rm{H}$ yields increase sharply from $\sqrt{s_{\rm{NN}}} = 27$ GeV to $4.5$ GeV, and appear to reach a maximum at $3$--$4$ GeV. Both the ${}^3_\Lambda \rm{H}$ yields and the ${}^3_\Lambda \rm{H}/\Lambda$ yields ratio are systematically lower than thermal model calculations by a factor of $\sim$2 in the reported energy range. The $\langle\pt\rangle$ of ${}^3_\Lambda \rm{H}$ also tends to be lower than Blast-Wave expectations based on kinetic freeze-out parameters extracted from light hadrons. These observations imply that the thermal model is inadequate to describe ${}^3_\Lambda \rm{H}$ production in the high baryon density region. In contrast, a hadronic transport model with coalescence afterburner, which implements a smaller coalescence probability for the loosely bound ${}^3_\Lambda \rm{H}$ versus the more tightly bound triton, qualitatively describes both the ${}^3_\Lambda \rm{H}$ yields and ${}^3_\Lambda \rm{H}/\Lambda$ yields ratio across the measured energy range. This success is also seen in describing the double ratio $({}^3_\Lambda {\rm H}/\Lambda)/(t/p)$, which is persistently suppressed ($\sim$0.4) relative to unity and exhibits no strong energy dependence.
Our data supports that the weaker hyperon–nucleon interaction, relative to the nucleon–nucleon interaction, has a direct impact on the formation probability of hypernuclei.
These results also offer a valuable reference to the measurement of heavier hypernuclei in future experiments.

We thank the RHIC Operations Group and SDCC at BNL, the NERSC Center at LBNL, and the Open Science Grid consortium for providing resources and support.  This work was supported in part by the Office of Nuclear Physics within the U.S. DOE Office of Science, the U.S. National Science Foundation, National Natural Science Foundation of China, Chinese Academy of Science, the Ministry of Science and Technology of China and the Chinese Ministry of Education, NSTC Taipei, the National Research Foundation of Korea, Czech Science Foundation and Ministry of Education, Youth and Sports of the Czech Republic, Hungarian National Research, Development and Innovation Office, New National Excellency Programme of the Hungarian Ministry of Human Capacities, Department of Atomic Energy and Department of Science and Technology of the Government of India, the National Science Centre and WUT ID-UB of Poland, German Bundesministerium f\"ur Bildung, Wissenschaft, Forschung and Technologie (BMBF), Helmholtz Association, Ministry of Education, Culture, Sports, Science, and Technology (MEXT), and Japan Society for the Promotion of Science (JSPS).

\normalem
\bibliographystyle{apsrev4-2}
\bibliography{ref.bib}

@article{STAR:2009sxc,
    author = "Abelev, B. I. and others",
    collaboration = "STAR",
    title = "{Identified particle production, azimuthal anisotropy, and interferometry measurements in Au+Au collisions at $\sNN$ = 9.2\,GeV}",
    eprint = "0909.4131",
    archivePrefix = "arXiv",
    primaryClass = "nucl-ex",
    doi = "10.1103/PhysRevC.81.024911",
    journal = "Phys. Rev. C",
    volume = "81",
    pages = "024911",
    year = "2010"
}

@article{Miller:2007ri,
    author = "Miller, Michael L. and Reygers, Klaus and Sanders, Stephen J. and Steinberg, Peter",
    title = "{Glauber modeling in high energy nuclear collisions}",
    eprint = "nucl-ex/0701025",
    archivePrefix = "arXiv",
    doi = "10.1146/annurev.nucl.57.090506.123020",
    journal = "Ann. Rev. Nucl. Part. Sci.",
    volume = "57",
    pages = "205--243",
    year = "2007"
}

@article{STAR:2021yiu,
    author = "Abdallah, M. S. and others",
    collaboration = "STAR",
    title = "{Disappearance of partonic collectivity in sNN=3GeV Au+Au collisions at RHIC}",
    eprint = "2108.00908",
    archivePrefix = "arXiv",
    primaryClass = "nucl-ex",
    doi = "10.1016/j.physletb.2022.137003",
    journal = "Phys. Lett. B",
    volume = "827",
    pages = "137003",
    year = "2022"
}

@article{Yang:2019bjr,
    author = "Yang, Qian",
    editor = "Antinori, Federico and Dainese, Andrea and Giubellino, Paolo and Greco, Vincenzo and Lombardo, Maria Paola and Scomparin, Enrico",
    collaboration = "STAR",
    title = "{The STAR BES-II and Forward Rapidity Physics and Upgrades}",
    doi = "10.1016/j.nuclphysa.2018.10.029",
    journal = "Nucl. Phys. A",
    volume = "982",
    pages = "951--954",
    year = "2019"
}

@article{Anderson:2003ur,
    author = "Anderson, M. and others",
    title = "{The Star time projection chamber: A Unique tool for studying high multiplicity events at RHIC}",
    eprint = "nucl-ex/0301015",
    archivePrefix = "arXiv",
    doi = "10.1016/S0168-9002(02)01964-2",
    journal = "Nucl. Instrum. Meth. A",
    volume = "499",
    pages = "659--678",
    year = "2003"
}

@article{WANG201790,
title = {Design and implementation of wire tension measurement system for MWPCs used in the STAR iTPC upgrade},
journal = {Nucl. Instrum. Meth. A},
volume = {859},
pages = {90-94},
year = {2017},
issn = {0168-9002},
doi = {https://doi.org/10.1016/j.nima.2017.04.005},
url = {https://www.sciencedirect.com/science/article/pii/S0168900217304254},
author = {Xu Wang and Fuwang Shen and Shuai Wang and Cunfeng Feng and Changyu Li and Peng Lu and Jim Thomas and Qinghua Xu and Chengguang Zhu}
}

@article{STAR:2022fnj,
    author = "Aboona, Bassam and others",
    collaboration = "STAR",
    title = "{Observation of Directed Flow of Hypernuclei H\ensuremath{\Lambda}3 and H\ensuremath{\Lambda}4 in sNN=3\,\,GeV Au+Au Collisions at RHIC}",
    eprint = "2211.16981",
    archivePrefix = "arXiv",
    primaryClass = "nucl-ex",
    doi = "10.1103/PhysRevLett.130.212301",
    journal = "Phys. Rev. Lett.",
    volume = "130",
    number = "21",
    pages = "212301",
    year = "2023"
}

@article{Ju:2023xvg,
    author = "Ju, Xin-Yue and others",
    title = "{Applying the Kalman filter particle method to strange and open charm hadron reconstruction in the STAR experiment}",
    doi = "10.1007/s41365-023-01320-1",
    journal = "Nucl. Sci. Tech.",
    volume = "34",
    number = "10",
    pages = "158",
    year = "2023"
}

@phdthesis{Zyzak:2016exl,
    author = "Zyzak, Maksym",
    title = "Online selection of short-lived particles on many-core computer architectures in the CBM experiment at FAIR",
    reportNumber = "GSI-2017-00683",
    school = "Frankfurt U.",
    year = "2016"
}

@article{STAR:2019bjj,
    author = "Adam, Jaroslav and others",
    collaboration = "STAR",
    title = "{Strange hadron production in Au+Au collisions at $\sqrt{s_{_{\mathrm{NN}}}}$ = 7.7, 11.5, 19.6, 27, and 39 GeV}",
    eprint = "1906.03732",
    archivePrefix = "arXiv",
    primaryClass = "nucl-ex",
    doi = "10.1103/PhysRevC.102.034909",
    journal = "Phys. Rev. C",
    volume = "102",
    number = "3",
    pages = "034909",
    year = "2020"
}

@article{STAR:2017sal,
    author = "Adamczyk, L. and others",
    collaboration = "STAR",
    title = "{Bulk Properties of the Medium Produced in Relativistic Heavy-Ion Collisions from the Beam Energy Scan Program}",
    eprint = "1701.07065",
    archivePrefix = "arXiv",
    primaryClass = "nucl-ex",
    doi = "10.1103/PhysRevC.96.044904",
    journal = "Phys. Rev. C",
    volume = "96",
    number = "4",
    pages = "044904",
    year = "2017"
}

@article{Vovchenko:2015idt,
    author = "Vovchenko, V. and Begun, V. V. and Gorenstein, M. I.",
    title = "{Hadron multiplicities and chemical freeze-out conditions in proton-proton and nucleus-nucleus collisions}",
    eprint = "1512.08025",
    archivePrefix = "arXiv",
    primaryClass = "nucl-th",
    doi = "10.1103/PhysRevC.93.064906",
    journal = "Phys. Rev. C",
    volume = "93",
    number = "6",
    pages = "064906",
    year = "2016"
}

@article{Reichert:2022mek,
    author = {Reichert, Tom and Steinheimer, Jan and Vovchenko, Volodymyr and D\"onigus, Benjamin and Bleicher, Marcus},
    title = "{Energy dependence of light hypernuclei production in heavy-ion collisions from a coalescence and statistical-thermal model perspective}",
    eprint = "2210.11876",
    archivePrefix = "arXiv",
    primaryClass = "nucl-th",
    doi = "10.1103/PhysRevC.107.014912",
    journal = "Phys. Rev. C",
    volume = "107",
    number = "1",
    pages = "014912",
    year = "2023"
}

@article{STAR:2022hbp,
    author = "Abdulhamid, Muhammad and others",
    collaboration = "STAR",
    title = "{Beam Energy Dependence of Triton Production and Yield Ratio ($\mathrm{N}_t \times \mathrm{N}_p/\mathrm{N}_d^2$) in Au+Au Collisions at RHIC}",
    eprint = "2209.08058",
    archivePrefix = "arXiv",
    primaryClass = "nucl-ex",
    doi = "10.1103/PhysRevLett.130.202301",
    journal = "Phys. Rev. Lett.",
    volume = "130",
    pages = "202301",
    year = "2023"
}

@article{STAR:2021orx,
    author = "Abdallah, Mohamed and others",
    collaboration = "STAR",
    title = "{Measurements of $H_\Lambda^3$ and $H_\Lambda^4$ Lifetimes and Yields in Au+Au Collisions in the High Baryon Density Region}",
    eprint = "2110.09513",
    archivePrefix = "arXiv",
    primaryClass = "nucl-ex",
    doi = "10.1103/PhysRevLett.128.202301",
    journal = "Phys. Rev. Lett.",
    volume = "128",
    number = "20",
    pages = "202301",
    year = "2022"
}

@article{Andronic:2010qu,
    author = "Andronic, A. and others",
    title = "{Production of light nuclei, hypernuclei and their antiparticles in relativistic nuclear collisions}",
    eprint = "1010.2995",
    archivePrefix = "arXiv",
    primaryClass = "nucl-th",
    doi = "10.1016/j.physletb.2011.01.053",
    journal = "Phys. Lett. B",
    volume = "697",
    pages = "203--207",
    year = "2011"
}

@article{Andronic:2017pug,
    author = "Andronic, Anton and Braun-Munzinger, Peter and Redlich, Krzysztof and Stachel, Johanna",
    title = "{Decoding the phase structure of QCD via particle production at high energy}",
    eprint = "1710.09425",
    archivePrefix = "arXiv",
    primaryClass = "nucl-th",
    doi = "10.1038/s41586-018-0491-6",
    journal = "Nature",
    volume = "561",
    number = "7723",
    pages = "321--330",
    year = "2018"
}

@article{Glassel:2021rod,
    author = {Gl\"a\ss{}el, Susanne and Kireyeu, Viktar and Voronyuk, Vadim and Aichelin, J\"org and Blume, Christoph and Bratkovskaya, Elena and Coci, Gabriele and Kolesnikov, Vadim and Winn, Michael},
    title = "{Cluster and hypercluster production in relativistic heavy-ion collisions within the parton-hadron-quantum-molecular-dynamics approach}",
    eprint = "2106.14839",
    archivePrefix = "arXiv",
    primaryClass = "nucl-th",
    doi = "10.1103/PhysRevC.105.014908",
    journal = "Phys. Rev. C",
    volume = "105",
    number = "1",
    pages = "014908",
    year = "2022"
}

@article{Bleicher:1999xi,
    author = "Bleicher, M. and others",
    title = "{Relativistic hadron hadron collisions in the ultrarelativistic quantum molecular dynamics model}",
    eprint = "hep-ph/9909407",
    archivePrefix = "arXiv",
    doi = "10.1088/0954-3899/25/9/308",
    journal = "J. Phys. G",
    volume = "25",
    pages = "1859--1896",
    year = "1999"
}

@article{Bass:1998ca,
    author = "Bass, S. A. and others",
    title = "{Microscopic models for ultrarelativistic heavy ion collisions}",
    eprint = "nucl-th/9803035",
    archivePrefix = "arXiv",
    doi = "10.1016/S0146-6410(98)00058-1",
    journal = "Prog. Part. Nucl. Phys.",
    volume = "41",
    pages = "255--369",
    year = "1998"
}

@article{Ji:2023fqp,
    author = "Ji, Yuanjing",
    collaboration = "STAR",
    title = "{Measurements on the production and properties of light hypernuclei at STAR}",
    doi = "10.1051/epjconf/202327604003",
    journal = "EPJ Web Conf.",
    volume = "276",
    pages = "04003",
    year = "2023"
}

@article{ALICE:2022sco,
    author = "Acharya, Shreyasi and others",
    collaboration = "ALICE",
    title = "{Measurement of the Lifetime and \ensuremath{\Lambda} Separation Energy of H\ensuremath{\Lambda}3}",
    eprint = "2209.07360",
    archivePrefix = "arXiv",
    primaryClass = "nucl-ex",
    reportNumber = "CERN-EP-2022-188",
    doi = "10.1103/PhysRevLett.131.102302",
    journal = "Phys. Rev. Lett.",
    volume = "131",
    number = "10",
    pages = "102302",
    year = "2023"
}

@article{ALICE:2015oer,
    author = "Adam, Jaroslav and others",
    collaboration = "ALICE",
    title = "{$^{3}_{\Lambda}\mathrm H$ and $^{3}_{\bar{\Lambda}} \overline{\mathrm H}$ production in Pb-Pb collisions at $\sqrt{s_{\rm NN}} =$ 2.76 TeV}",
    eprint = "1506.08453",
    archivePrefix = "arXiv",
    primaryClass = "nucl-ex",
    reportNumber = "CERN-PH-EP-2015-105",
    doi = "10.1016/j.physletb.2016.01.040",
    journal = "Phys. Lett. B",
    volume = "754",
    pages = "360--372",
    year = "2016"
}

@article{ALICE:2019vlx,
    author = "Acharya, Shreyasi and others",
    collaboration = "ALICE",
    title = "{$^3_\Lambda\mathrm{H}$ and $^3_{\bar{\Lambda}}\mathrm{\overline{H}}$ lifetime measurement in Pb-Pb collisions at $\sqrt{s_{\mathrm{NN}}} = $ 5.02 TeV via two-body decay}",
    eprint = "1907.06906",
    archivePrefix = "arXiv",
    primaryClass = "nucl-ex",
    reportNumber = "CERN-EP-2019-148",
    doi = "10.1016/j.physletb.2019.134905",
    journal = "Phys. Lett. B",
    volume = "797",
    pages = "134905",
    year = "2019"
}

@article{STAR:2010gyg,
    author = "Abelev, B. I. and others",
    collaboration = "STAR",
    title = "{Observation of an Antimatter Hypernucleus}",
    eprint = "1003.2030",
    archivePrefix = "arXiv",
    primaryClass = "nucl-ex",
    doi = "10.1126/science.1183980",
    journal = "Science",
    volume = "328",
    pages = "58--62",
    year = "2010"
}

@inproceedings{Syresin:2019vzo,
    author = "Syresin, Evgeny and others",
    title = "{NICA Accelerator Complex at JINR}",
    booktitle = "{10th International Particle Accelerator Conference}",
    doi = "10.18429/JACoW-IPAC2019-MOPMP014",
    pages = "MOPMP014",
    year = "2019",
    collaboration = "NICA"
}

@article{Herrmann:2022jkv,
    author = "Herrmann, Norbert",
    collaboration = "CBM",
    title = "{Status and Perspectives of the CBM experiment at FAIR}",
    doi = "10.1051/epjconf/202225909001",
    journal = "EPJ Web Conf.",
    volume = "259",
    pages = "09001",
    year = "2022"
}

@article{RAPPOLD2015129,
title = {Hypernuclear production cross section in the reaction of 6Li + 12C at 2 A GeV},
journal = {Physics Letters B},
volume = {747},
pages = {129-134},
year = {2015},
issn = {0370-2693},
doi = {https://doi.org/10.1016/j.physletb.2015.05.059},
url = {https://www.sciencedirect.com/science/article/pii/S0370269315003962},
author = {C. Rappold and others},
collaboration = "HypHI"
}

@article{ALICE:2021puh,
    author = "Acharya, Shreyasi and others",
    collaboration = "ALICE",
    title = "{Hypertriton Production in p-Pb Collisions at $\sqrt {s_{NN}}$=5.02\,\,TeV}",
    eprint = "2107.10627",
    archivePrefix = "arXiv",
    primaryClass = "nucl-ex",
    reportNumber = "CERN-EP-2021-139",
    doi = "10.1103/PhysRevLett.128.252003",
    journal = "Phys. Rev. Lett.",
    volume = "128",
    number = "25",
    pages = "252003",
    year = "2022"
}

@misc{Aoki:2021cqa,
  author = "Aoki, Kazuya and others",
  title = "{Extension of the J-PARC Hadron Experimental Facility: Third White Paper}",
  eprint = "2110.04462",
  archivePrefix = "arXiv",
  primaryClass = "nucl-ex",
  year = "2021"
}

@article{Zhou:2022pxl,
    author = "Zhou, Xiaohong and Yang, Jiancheng",
    collaboration = "HIAF project Team",
    title = "{Status of the high-intensity heavy-ion accelerator facility in China}",
    doi = "10.1007/s43673-022-00064-1",
    journal = "AAPPS Bull.",
    volume = "32",
    number = "1",
    pages = "35",
    year = "2022"
}

@article{STAR:2023fbc,
    author = "Abdulhamid, Muhammad and others",
    collaboration = "STAR",
    title = "{Observation of the antimatter hypernucleus ${}_{\bar{{\boldsymbol{\Lambda }}}}{}^{{\bf{4}}}\bar{{\bf{H}}}$}",
    eprint = "2310.12674",
    archivePrefix = "arXiv",
    primaryClass = "nucl-ex",
    doi = "10.1038/s41586-024-07823-0",
    journal = "Nature",
    volume = "632",
    number = "8027",
    pages = "1026--1031",
    year = "2024"
}

@article{Llope:2014nva,
    author = "Llope, W. J. and others",
    title = "{The STAR Vertex Position Detector}",
    eprint = "1403.6855",
    archivePrefix = "arXiv",
    primaryClass = "physics.ins-det",
    doi = "10.1016/j.nima.2014.04.080",
    journal = "Nucl. Instrum. Meth. A",
    volume = "759",
    pages = "23--28",
    year = "2014"
}

@article{Xu:2016alq,
    author = "Xu, Yi-Fei and Chen, Jin-Hui and Ma, Yu-Gang and Tang, Ai-Hong and Xu, Zhang-Bu and Zhu, Yu-Hui",
    title = "{Physics performance of the STAR zero degree calorimeter at relativistic heavy ion collider}",
    doi = "10.1007/s41365-016-0129-z",
    journal = "Nucl. Sci. Tech.",
    volume = "27",
    number = "6",
    pages = "126",
    year = "2016"
}

@article{BONNER2003181,
title = {A single Time-of-Flight tray based on multigap resistive plate chambers for the STAR experiment at RHIC},
journal = {Nucl. Instrum. Meth. A},
volume = {508},
number = {1},
pages = {181-184},
year = {2003},
note = {Proceedings of the Sixth International Workshop on Resistive Plate Chambers and Related Detectors},
issn = {0168-9002},
doi = {https://doi.org/10.1016/S0168-9002(03)01347-0},
author = {B Bonner and H Chen and G Eppley and F Geurts and J Lamas-Valverde and Ch Li and W.J Llope and T Nussbaum and E Platner and J Roberts}
}

@article{Whitten:2008zz,
    author = "Whitten, C. A.",
    editor = "Kponou, Ahovi and Makdisi, Yousef and Zelenski, Anatoli",
    collaboration = "STAR",
    title = "{The beam-beam counter: A local polarimeter at STAR}",
    doi = "10.1063/1.2888113",
    journal = "AIP Conf. Proc.",
    volume = "980",
    number = "1",
    pages = "390--396",
    year = "2008"
}

@article{Eckert:2022dyz,
    author = "Eckert, Philipp and others",
    title = "{Systematic treatment of hypernuclear data and application to the hypertriton}",
    eprint = "2201.02368",
    archivePrefix = "arXiv",
    primaryClass = "physics.data-an",
    doi = "10.31349/SuplRevMexFis.3.0308069",
    journal = "Rev. Mex. Fis. Suppl.",
    volume = "3",
    number = "3",
    pages = "0308069",
    year = "2022"
}

@article{STAR:2024znc,
    author = "Abdulhamid, M. I. and others",
    collaboration = "STAR",
    title = "{Strangeness production in $ \sqrt{s_{\textrm{NN}}} $ = 3 GeV Au+Au collisions at RHIC}",
    eprint = "2407.10110",
    archivePrefix = "arXiv",
    primaryClass = "nucl-ex",
    doi = "10.1007/JHEP10(2024)139",
    journal = "J. High Energ. Phys.",
    volume = "10",
    pages = "139",
    year = "2024"
}

@article{STAR:2023uxk,
    author = "Abdulhamid, Muhammad and others",
    collaboration = "STAR",
    title = "{Production of protons and light nuclei in Au+Au collisions at sNN=3 GeV with the STAR detector}",
    eprint = "2311.11020",
    archivePrefix = "arXiv",
    primaryClass = "nucl-ex",
    doi = "10.1103/PhysRevC.110.054911",
    journal = "Phys. Rev. C",
    volume = "110",
    number = "5",
    pages = "054911",
    year = "2024"
}

@article{Sun:2022xjr,
    author = "Sun, Kai-Jia and Wang, Rui and Ko, Che Ming and Ma, Yu-Gang and Shen, Chun",
    title = "{Unveiling the dynamics of little-bang nucleosynthesis}",
    eprint = "2207.12532",
    archivePrefix = "arXiv",
    primaryClass = "nucl-th",
    doi = "10.1038/s41467-024-45474-x",
    journal = "Nature Commun.",
    volume = "15",
    number = "1",
    pages = "1074",
    year = "2024"
}

@article{ALICE:2022veq,
    author = "Acharya, Shreyasi and others",
    collaboration = "ALICE",
    title = "{Light (anti)nuclei production in Pb-Pb collisions at sNN=5.02~TeV}",
    eprint = "2211.14015",
    archivePrefix = "arXiv",
    primaryClass = "nucl-ex",
    reportNumber = "CERN-EP-2022-263",
    doi = "10.1103/PhysRevC.107.064904",
    journal = "Phys. Rev. C",
    volume = "107",
    number = "6",
    pages = "064904",
    year = "2023"
}

@article{STAR:2017gxa,
    author = "Adamczyk, L. and others",
    collaboration = "STAR",
    title = "{Measurement of the $^3_{\Lambda}$H lifetime in Au+Au collisions at the BNL Relativistic Heavy Ion Collider}",
    eprint = "1710.00436",
    archivePrefix = "arXiv",
    primaryClass = "nucl-ex",
    doi = "10.1103/PhysRevC.97.054909",
    journal = "Phys. Rev. C",
    volume = "97",
    number = "5",
    pages = "054909",
    year = "2018"
}

@article{STAR:2019wjm,
    author = "Adam, J. and others",
    collaboration = "STAR",
    title = "{Measurement of the mass difference and the binding energy of the hypertriton and antihypertriton}",
    eprint = "1904.10520",
    archivePrefix = "arXiv",
    primaryClass = "hep-ex",
    doi = "10.1038/s41567-020-0799-7",
    journal = "Nature Phys.",
    volume = "16",
    number = "4",
    pages = "409--412",
    year = "2020"
}

@article{STAR:2019sjh,
    author = "Adam, Jaroslav and others",
    collaboration = "STAR",
    title = "{Beam energy dependence of (anti-)deuteron production in Au + Au collisions at the BNL Relativistic Heavy Ion Collider}",
    eprint = "1903.11778",
    archivePrefix = "arXiv",
    primaryClass = "nucl-ex",
    doi = "10.1103/PhysRevC.99.064905",
    journal = "Phys. Rev. C",
    volume = "99",
    number = "6",
    pages = "064905",
    year = "2019"
}

@article{E802:1999hit,
    author = "Ahle, L. and others",
    collaboration = "E802",
    title = "{Proton and deuteron production in Au + Au reactions at 11.6/A-GeV/c}",
    doi = "10.1103/PhysRevC.60.064901",
    journal = "Phys. Rev. C",
    volume = "60",
    pages = "064901",
    year = "1999"
}

@article{Chen:2023mel,
    author = "Chen, Jinhui and Dong, Xin and Ma, Yu-Gang and Xu, Zhangbu",
    title = "{Measurements of the lightest hypernucleus (H\ensuremath{\Lambda}3): progress and perspective}",
    eprint = "2311.09877",
    archivePrefix = "arXiv",
    primaryClass = "nucl-ex",
    doi = "10.1016/j.scib.2023.11.045",
    journal = "Sci. Bull.",
    volume = "68",
    pages = "3252--3260",
    year = "2023"
}

@article{Sun:2018mqq,
    author = {Sun, Kai-Jia and Ko, Che Ming and D\"onigus, Benjamin},
    title = "{Suppression of light nuclei production in collisions of small systems at the Large Hadron Collider}",
    eprint = "1812.05175",
    archivePrefix = "arXiv",
    primaryClass = "nucl-th",
    doi = "10.1016/j.physletb.2019.03.033",
    journal = "Phys. Lett. B",
    volume = "792",
    pages = "132--137",
    year = "2019"
}

@article{Dong:2018cye,
    author = "Dong, Zi-Jian and Chen, Gang and Wang, Quan-Yu and She, Zhi-Lei and Yan, Yu-Liang and Liu, Feng-Xian and Zhou, Dai-Mei and Sa, Ben-Hao",
    title = "{Energy dependence of light (anti)nuclei and (anti)hypertriton production in the Au-Au collision from $\sqrt{s_{NN}} = 11.5$ to 5020 GeV}",
    eprint = "1803.01547",
    archivePrefix = "arXiv",
    primaryClass = "nucl-th",
    doi = "10.1140/epja/i2018-12580-8",
    journal = "Eur. Phys. J. A",
    volume = "54",
    number = "9",
    pages = "144",
    year = "2018"
}

@article{Vovchenko:2020dmv,
    author = {Vovchenko, Volodymyr and D\"onigus, Benjamin and Kardan, Behruz and Lorenz, Manuel and Stoecker, Horst},
    title = "{Feeddown contributions from unstable nuclei in relativistic heavy-ion collisions}",
    eprint = "2004.04411",
    archivePrefix = "arXiv",
    primaryClass = "nucl-th",
    doi = "10.1016/j.physletb.2020.135746",
    journal = "Phys. Lett.",
    volume = "B",
    pages = "135746",
    year = "2020"
}

@misc{Mahlein:2025bla,
    author = "Mahlein, Maximilian and Singh, Bhawani and Viviani, Michele and Bellini, Francesca and Fabbietti, Laura and Kievsky, Alejandro and Marcucci, Laura Elisa",
    title = "{ToMCCA-3: A realistic 3-body coalescence model}",
    eprint = "2504.02491",
    archivePrefix = "arXiv",
    primaryClass = "hep-ph",
    reportNumber = "JLAB-PHY-25-4267",
    year = "2025"
}

@article{Bellini:2020cbj,
    author = "Bellini, Francesca and Blum, Kfir and Kalweit, Alexander Phillip and Puccio, Maximiliano",
    title = "{Examination of coalescence as the origin of nuclei in hadronic collisions}",
    eprint = "2007.01750",
    archivePrefix = "arXiv",
    primaryClass = "nucl-th",
    doi = "10.1103/PhysRevC.103.014907",
    journal = "Phys. Rev. C",
    volume = "103",
    number = "1",
    pages = "014907",
    year = "2021"
}

@inproceedings{Barlow:2002yb,
    author = "Barlow, Roger",
    title = "{Systematic errors: Facts and fictions}",
    booktitle = "{Conference on Advanced Statistical Techniques in Particle Physics}",
    eprint = "hep-ex/0207026",
    archivePrefix = "arXiv",
    reportNumber = "MAN-HEP-02-01",
    pages = "134--144",
    month = "7",
    year = "2002"
}

@article{Tolos:2020aln,
    author = "Tolos, Laura and Fabbietti, Laura",
    title = "{Strangeness in Nuclei and Neutron Stars}",
    eprint = "2002.09223",
    archivePrefix = "arXiv",
    primaryClass = "nucl-ex",
    doi = "10.1016/j.ppnp.2020.103770",
    journal = "Prog. Part. Nucl. Phys.",
    volume = "112",
    pages = "103770",
    year = "2020"
}

@article{Burgio:2021vgk,
    author = "Burgio, G. F. and Schulze, H. -J. and Vidana, I. and Wei, J. -B.",
    title = "{Neutron stars and the nuclear equation of state}",
    eprint = "2105.03747",
    archivePrefix = "arXiv",
    primaryClass = "nucl-th",
    doi = "10.1016/j.ppnp.2021.103879",
    journal = "Prog. Part. Nucl. Phys.",
    volume = "120",
    pages = "103879",
    year = "2021"
}

@article{Gal:2016boi,
    author = "Gal, A. and Hungerford, E. V. and Millener, D. J.",
    title = "{Strangeness in nuclear physics}",
    eprint = "1605.00557",
    archivePrefix = "arXiv",
    primaryClass = "nucl-th",
    doi = "10.1103/RevModPhys.88.035004",
    journal = "Rev. Mod. Phys.",
    volume = "88",
    number = "3",
    pages = "035004",
    year = "2016"
}

@article{Gerstung:2020ktv,
    author = "Gerstung, Dominik and Kaiser, Norbert and Weise, Wolfram",
    title = "{Hyperon{\textendash}nucleon three-body forces and strangeness in neutron stars}",
    eprint = "2001.10563",
    archivePrefix = "arXiv",
    primaryClass = "nucl-th",
    doi = "10.1140/epja/s10050-020-00180-2",
    journal = "Eur. Phys. J. A",
    volume = "56",
    number = "6",
    pages = "175",
    year = "2020"
}

@article{Lonardoni:2014bwa,
    author = "Lonardoni, Diego and Lovato, Alessandro and Gandolfi, Stefano and Pederiva, Francesco",
    title = "{Hyperon Puzzle: Hints from Quantum Monte Carlo Calculations}",
    eprint = "1407.4448",
    archivePrefix = "arXiv",
    primaryClass = "nucl-th",
    reportNumber = "LA-UR-14-25265",
    doi = "10.1103/PhysRevLett.114.092301",
    journal = "Phys. Rev. Lett.",
    volume = "114",
    number = "9",
    pages = "092301",
    year = "2015"
}

@techreport{Brun:1987ma,
    author = "Brun, R. and others",
    title = "{GEANT3}",
    reportNumber = "CERN-DD-EE-84-1",
    year = "1987"
}

@article{STAR:2014clz,
    author = "Adamczyk, L. and others",
    collaboration = "STAR",
    title = "{Beam-Energy Dependence of the Directed Flow of Protons, Antiprotons, and Pions in Au+Au Collisions}",
    eprint = "1401.3043",
    archivePrefix = "arXiv",
    primaryClass = "nucl-ex",
    doi = "10.1103/PhysRevLett.112.162301",
    journal = "Phys. Rev. Lett.",
    volume = "112",
    number = "16",
    pages = "162301",
    year = "2014"
}

@article{Schnedermann:1993ws,
    author = "Schnedermann, Ekkard and Sollfrank, Josef and Heinz, Ulrich W.",
    title = "{Thermal phenomenology of hadrons from 200-A/GeV S+S collisions}",
    eprint = "nucl-th/9307020",
    archivePrefix = "arXiv",
    reportNumber = "TPR-93-16",
    doi = "10.1103/PhysRevC.48.2462",
    journal = "Phys. Rev. C",
    volume = "48",
    pages = "2462--2475",
    year = "1993"
}

@inbook{Braun-Munzinger:2003pwq,
    title="Particle production in heavy ion collisions",
    author = "Braun-Munzinger, Peter and Redlich, Krzysztof and Stachel, Johanna",
    booktitle = {Quark–Gluon Plasma 3},
    chapter = {},
    pages = {491-599},
    doi = {10.1142/9789812795533_0008},
    URL = {https://www.worldscientific.com/doi/abs/10.1142/9789812795533_0008},
    eprint = "nucl-th/0304013",
    archivePrefix = "arXiv",
    month = "4",
    year = "2003"
}

@article{Steinheimer:2012tb,
    author = "Steinheimer, J. and Gudima, K. and Botvina, A. and Mishustin, I. and Bleicher, M. and Stocker, H.",
    title = "{Hypernuclei, dibaryon and antinuclei production in high energy heavy ion collisions: Thermal production versus Coalescence}",
    eprint = "1203.2547",
    archivePrefix = "arXiv",
    primaryClass = "nucl-th",
    doi = "10.1016/j.physletb.2012.06.069",
    journal = "Phys. Lett. B",
    volume = "714",
    pages = "85--91",
    year = "2012"
}

@article{Judd:2018zbg,
    author = "Judd, E. G. and others",
    title = "{The evolution of the STAR Trigger System}",
    doi = "10.1016/j.nima.2018.03.070",
    journal = "Nucl. Instrum. Meth. A",
    volume = "902",
    pages = "228--237",
    year = "2018"
}

@article{STAR:2021hyx,
    author = "Abdallah, M. S. and others",
    collaboration = "STAR",
    title = "{Probing strangeness canonical ensemble with K{\ensuremath{-}}, {\ensuremath{\phi}}(1020) and {\ensuremath{\Xi}}{\ensuremath{-}} production in Au+Au collisions at sNN=3 GeV}",
    eprint = "2108.00924",
    archivePrefix = "arXiv",
    primaryClass = "nucl-ex",
    doi = "10.1016/j.physletb.2022.137152",
    journal = "Phys. Lett. B",
    volume = "831",
    pages = "137152",
    year = "2022"
}

@article{Kamada:1997rv,
    author = "Kamada, H. and Golak, J. and Miyagawa, K. and Witala, H. and Gloeckle, Walter",
    title = "{Pi mesonic decay of the hypertriton}",
    eprint = "nucl-th/9709035",
    archivePrefix = "arXiv",
    doi = "10.1103/PhysRevC.57.1595",
    journal = "Phys. Rev. C",
    volume = "57",
    pages = "1595--1603",
    year = "1998"
}

@article{Gazda:2023fow,
    author = "Gazda, D. and P\'erez-Obiol, A. and Friedman, E. and Gal, A.",
    collaboration = "ALICE",
    title = "{Hypertriton lifetime}",
    eprint = "2310.03087",
    archivePrefix = "arXiv",
    primaryClass = "nucl-th",
    doi = "10.1103/PhysRevC.109.024001",
    journal = "Phys. Rev. C",
    volume = "109",
    number = "2",
    pages = "024001",
    year = "2024"
}

@article{Golak:1996hj,
    author = "Golak, J. and Miyagawa, K. and Kamada, H. and Witala, H. and Gloeckle, Walter and Parreno, A. and Ramos, A. and Bennhold, C.",
    title = "{The Non-mesonic weak decay of the hypertriton}",
    eprint = "nucl-th/9612065",
    archivePrefix = "arXiv",
    doi = "10.1103/PhysRevC.56.2892",
    journal = "Phys. Rev. C",
    volume = "55",
    pages = "2196--2213",
    year = "1997",
    note = "[Erratum: Phys.Rev.C 56, 2892--2892 (1997)]"
}

@article{Vovchenko:2019pjl,
    author = "Vovchenko, Volodymyr and Stoecker, Horst",
    title = "{Thermal-FIST: A package for heavy-ion collisions and hadronic equation of state}",
    eprint = "1901.05249",
    archivePrefix = "arXiv",
    primaryClass = "nucl-th",
    doi = "10.1016/j.cpc.2019.06.024",
    journal = "Comput. Phys. Commun.",
    volume = "244",
    pages = "295--310",
    year = "2019"
}

@article{FOPI:2006ifg,
    author = "Reisdorf, W. and others",
    collaboration = "FOPI",
    title = "{Systematics of pion emission in heavy ion collisions in the 1A- GeV regime}",
    eprint = "nucl-ex/0610025",
    archivePrefix = "arXiv",
    doi = "10.1016/j.nuclphysa.2006.10.085",
    journal = "Nucl. Phys. A",
    volume = "781",
    pages = "459--508",
    year = "2007"
}

@article{E864:2000auv,
    author = "Armstrong, T. A. and others",
    collaboration = "E864",
    title = "{Measurements of light nuclei production in 11.5-A-GeV/c Au + Pb heavy ion collisions}",
    eprint = "nucl-ex/0003009",
    archivePrefix = "arXiv",
    doi = "10.1103/PhysRevC.61.064908",
    journal = "Phys. Rev. C",
    volume = "61",
    pages = "064908",
    year = "2000"
}

@article{PHENIX:2004vqi,
    author = "Adler, S. S. and others",
    collaboration = "PHENIX",
    title = "{Deuteron and antideuteron production in Au + Au collisions at s(NN)**(1/2) = 200-GeV}",
    eprint = "nucl-ex/0406004",
    archivePrefix = "arXiv",
    doi = "10.1103/PhysRevLett.94.122302",
    journal = "Phys. Rev. Lett.",
    volume = "94",
    pages = "122302",
    year = "2005"
}

@article{PHENIX:2003iij,
    author = "Adler, S. S. and others",
    collaboration = "PHENIX",
    title = "{Identified charged particle spectra and yields in Au+Au collisions at S(NN)**1/2 = 200-GeV}",
    eprint = "nucl-ex/0307022",
    archivePrefix = "arXiv",
    doi = "10.1103/PhysRevC.69.034909",
    journal = "Phys. Rev. C",
    volume = "69",
    pages = "034909",
    year = "2004"
}

@article{ALICE:2015wav,
    author = "Adam, Jaroslav and others",
    collaboration = "ALICE",
    title = "{Production of light nuclei and anti-nuclei in pp and Pb-Pb collisions at energies available at the CERN Large Hadron Collider}",
    eprint = "1506.08951",
    archivePrefix = "arXiv",
    primaryClass = "nucl-ex",
    reportNumber = "CERN-PH-EP-2015-025",
    doi = "10.1103/PhysRevC.93.024917",
    journal = "Phys. Rev. C",
    volume = "93",
    number = "2",
    pages = "024917",
    year = "2016"
}

\end{document}


\title{End Matter: Collision Energy Dependence of Hypertriton Production in Au+Au Collisions at RHIC}
\author{The STAR Collaboration}
\date{\today}
\maketitle

\section{Centrality Dependence of Hypertriton Production}
\label{sec:1040cent_results}

As a complement to the spectra in 0-10\% centrality shown in the main text of the paper, Fig. \ref{fig:ptspectra1040} presents the transverse momentum spectra ($p_{\rm T}$) of $\rm {}^{3}_{\Lambda}H$ at mid-rapidity in 10-40\% mid-central Au+Au collisions at $\sNN$ = 3.2--27\,GeV.

\begin{figure}[htb]
\includegraphics[width=0.85\linewidth]{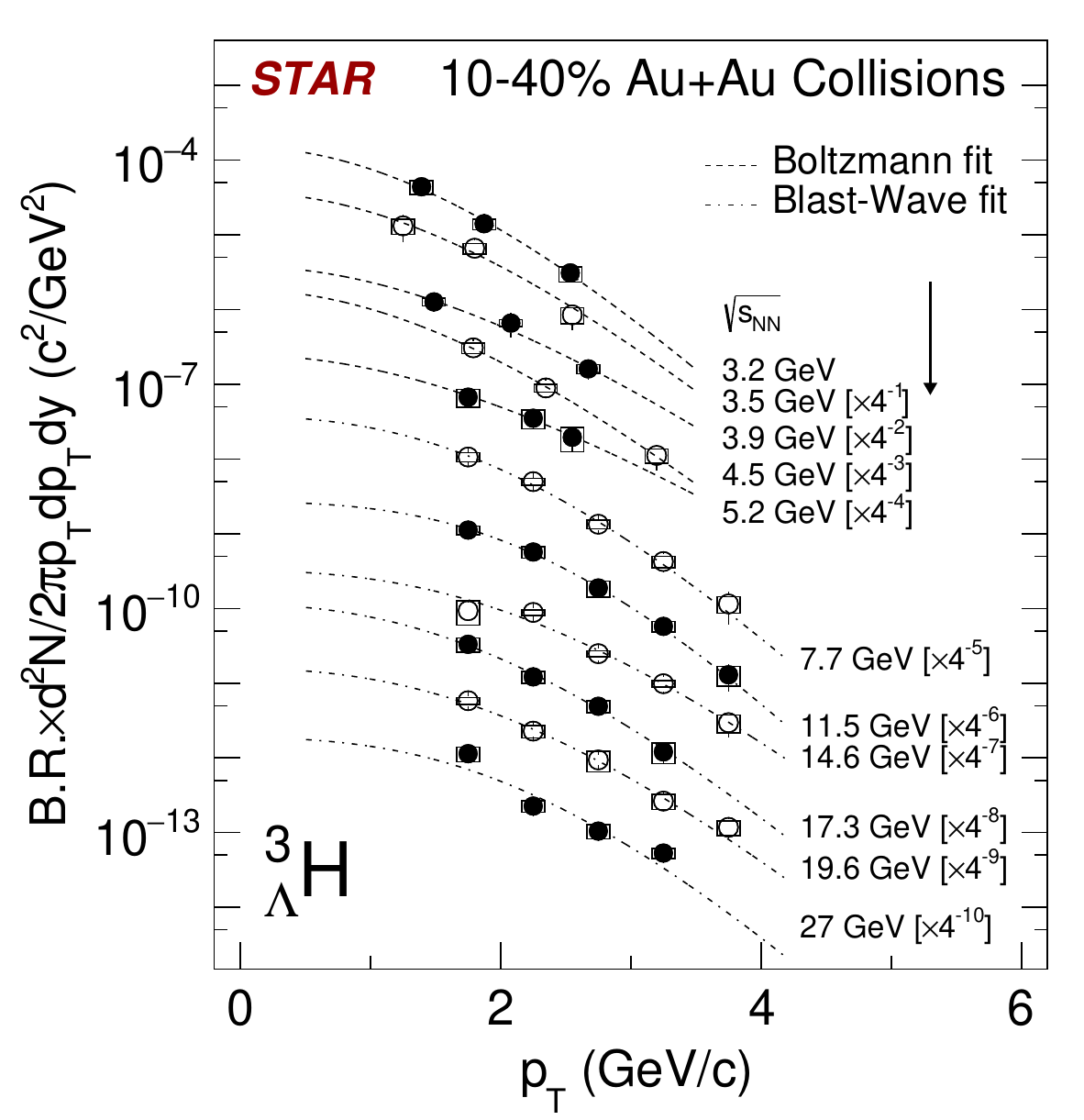}
\caption{\label{fig:ptspectra1040} The transverse momentum spectra of $\rm {}^{3}_{\Lambda}H$ from $\sNN=$ 3.2 to 27 GeV in Au+Au collisions at 10-40\% centrality at mid-rapidity. The spectra are measured in the rapidity ranges of $-0.5<y<0$ for $\sNN = 3.2$–$4.5$ GeV, $-0.8<y<-0.3$ for $\sNN = 5.2$ GeV, and $-0.5<y<0.5$ for $\sNN = 7.7$–$27$ GeV. The dashed and dot-dashed lines are the Boltzmann and Blast-Wave fits to the data, respectively. The boxes indicate systematic uncertainties, while the vertical lines represent statistical uncertainties.}
\label{fig:ptspectra1040}
\end{figure}

\section{Systematic uncertainties}
\label{sec:systematic_source}
Table \ref{tab:sys_dndy} and \ref{tab:sys_mpt} list the relative systematic uncertainties of $\rm {}^{3}_{\Lambda}H$ $dN/dy$ and $\langle p_{\rm T} \rangle$ at mid-rapidity from different sources for 0-10\% centrality at $\snn$ = 3.2--5.2 and 7.7--27 GeV. The rapidity range is the same as Fig. 3 and Fig. 4 in the main text of the paper.

\begin{table}
\centering
\begin{tabular}{ c | c| c}
\hline
 \multicolumn{3}{c}{$dN/dy$ in Au+Au collisions at 0-10\% centrality} \\
  \hline
 $\snn$ (GeV) & 3.2--5.2 & 7.7--27 \\
 \hline
 Raw count extraction & $1-6\%$ & $3-6\%$ \\
 Tracking efficiency & $10-11\%$ & $5-12\%$ \\
 $\hyt$ lifetime & $<1-5\%$ & $6-13\%$ \\
 Topological cuts & $5-17\%$ & $9-19\%$\\
 Extrapolation & $5-11\%$ & $12-28\%$ \\
 \hline
 Total & $15-21\%$ & $18-35\%$ \\
\hline
\end{tabular}
\caption{The relative systematic uncertainties of $dN/dy$ in Au+Au collisions at 0-10\% centrality. The rapidity range is $-0.5<y<0$ for $\snn=$ 3.2--4.5 GeV, $-0.8<y<-0.3$ for $\snn=$ 5.2 GeV, and $|y|<0.5$ for $\snn=$ 7.7--27 GeV.}
\label{tab:sys_dndy}
\end{table}

\begin{table}
\centering
\begin{tabular}{ c | c| c}
\hline
 \multicolumn{3}{c}{$\langle p_{\rm T} \rangle$ in Au+Au collisions at 0-10\% centrality} \\
  \hline
 $\snn$ (GeV) & 3.2 - 5.2 & 7.7 - 27 \\
 \hline
 Raw count extraction & $<1-4\%$ & $<1-3\%$ \\
 Tracking efficiency & $<1\%$ & $<1-5\%$ \\
 $\hyt$ lifetime & $<1-1\%$ & $<1-3\%$ \\
 Topological cuts & $2-7\%$ & $3-10\%$\\
 Extrapolation & $1-5\%$ & $5-9\%$ \\
 \hline
 Total & $2-8\%$ & $7-15\%$ \\
\hline
\end{tabular}
\caption{The relative systematic uncertainties of $\langle p_{T} \rangle$ in Au+Au collisions at 0-10\% centrality. The rapidity range is $-0.5<y<0$ for $\snn=$ 3.2--4.5 GeV, $-0.8<y<-0.3$ for $\snn=$ 5.2 GeV, and $|y|<0.5$ for $\snn=$ 7.7--27 GeV.}
\label{tab:sys_mpt}
\end{table}

\section{Thermal Model and Blast-Wave Calculations}
\label{sec:thermal}

For our thermal model calculations, we use the Thermal-FIST package~\cite{Vovchenko:2019pjl}. Decays from unstable nuclei are included in the calculations. The collision energy dependence of
the temperature and baryochemical potential, $T(\sqrt{s_{\rm{NN}}})$ and $\mu_{B}(\sqrt{s_{\rm{NN}}})$, is parameterized using the chemical freeze-out curve from Ref.~\cite{Vovchenko:2015idt}, which is constrained from global data. The strangeness canonical ensemble is used with a strangeness correlation length of 3.55 fm, which best describes the $\Lambda/p$ ratio at $\sqrt{s_{\rm{NN}}}=3$ GeV~\cite{STAR:2024znc}. $\gamma_S$ is set to 1. The effective chemical freeze-out radii $R(\sqrt{s_{\rm{NN}}})$ for $0$-$10\%$ Au+Au collisions are fixed using the same method as in Ref.~\cite{Reichert:2022mek}: We use STAR mid-rapidity charged pion multiplicity data~\cite{STAR:2017sal} to parameterize its
collision energy dependence from 3 GeV to 27 GeV. The following function is used to parameterize the pion multiplicity:
\begin{align}\label{eq:pion}
\frac{dN_\pi^+}{dy} + \frac{dN_\pi^-}{dy} = a \times s_{\rm{NN}}^b \times \ln (s_{\rm{NN}}) - c,
\end{align}
\noindent At each $\sqrt{s_{\rm{NN}}}$, we set the effective freeze-out volume $V(\sqrt{s_{\rm{NN}}})=\frac{4\pi}{3}R(\sqrt{s_{\rm{NN}}})^3$ to a value such that the thermal model reproduces pion multiplicity from Eq.~\ref{eq:pion}. Table~\ref{tab:chemicalparameters} lists the chemical freeze-out parameters used in this manuscript. We have verified that these calculations well reproduce the $p$~\cite{STAR:2023uxk, STAR:2017sal}  and $\Lambda$~\cite{STAR:2024znc, STAR:2019bjj} yields from  $\sqrt{s_{\rm{NN}}}=3$ to 27 GeV. 

The Blast-Wave model assumes that particles are emitted from a locally thermalized source at the kinetic freeze-out temperature and move with a common collective transverse flow velocity~\cite{Schnedermann:1993ws,STAR:2017sal}. 
In this framework, the invariant $p_{\rm T}$ spectra of the particles can be expressed as:
\begin{equation}
\begin{split}
 \frac{1}{2\pi p_{\mathrm{T}}}\frac{d^{2}N}{dp_{\mathrm{T}}dy}
 \propto
 \int_{0}^{R} r\,dr\,
m_{\mathrm{T}}\,
I_{0}\!\left(\frac{p_{\mathrm{T}}\sinh\rho(r)}{T_{\mathrm{kin}}}\right) \\
\times
K_{1}\!\left(\frac{m_{\mathrm{T}}\cosh\rho(r)}{T_{\mathrm{kin}}}\right),
\end{split}
\label{eq:bw_spectrum}
\end{equation}
where $m_{\mathrm{T}}$ is the transverse mass of the particle, $I_{0}$ and $K_{1}$ are modified Bessel functions, $T_{\rm{kin}}$ is the kinetic freeze-out temperature, and $\rho(r) = \tanh^{-1}\beta$
is the velocity profile. The transverse radial flow velocity $\beta$ in the region $0\leq r\leq R$ can be expressed as $\beta = \beta_{S}(r/R)^n$, where $\beta_{S}$ is the surface velocity, and $r/R$ is the relative radial position in the thermal source, and the exponent $n$ reflects the form of the flow velocity profile. Average transverse radial flow
velocity $\beta$ can then be obtained from $\langle \beta \rangle=\frac{2}{2+n}\beta_{\mathrm{S}}$.

For Blast-Wave calculations in Fig. 3 of the main text of the paper, the freeze-out parameters including the kinetic freezeout temperature $T_{\rm kin}$, average radial velocity $\langle \beta \rangle,$ and radial flow profile parameter $n$ are obtained from published fits to the measured light hadron ($\pi$,$K$,$p$) spectra~\cite{STAR:2017sal, STAR:2023uxk, E802:1999hit}. Table~\ref{tab:kineticparameters} lists the kinetic freeze-out parameters used in this manuscript. 

\begin{table}[h]
    \centering
    \begin{tabular}{c c c c}
        \toprule
        $\sqrt{s_{\rm{NN}}}$ (GeV) & $T_{\rm ch}$ (MeV) & $\mu_{B}$ (MeV) & R (fm) \\
        \midrule
        3     & 85.1  & 727.9  & 8.28  \\
        3.2   & 91.3  & 704.1  & 7.87  \\
        3.5   & 99.1  & 671.2  & 7.41  \\
        3.9   & 107.6 & 631.8  & 6.99  \\
        4.5   & 117.2 & 580.7  & 6.62  \\
        7.7   & 140.2 & 405.7  & 6.15  \\
        11.5  & 148.5 & 298.7  & 6.22  \\
        14.6  & 151.4 & 245.9  & 6.34  \\
        19.6  & 153.7 & 191.3  & 6.51  \\
        27    & 155.2 & 143.9  & 6.72  \\
        \bottomrule
    \end{tabular}
    \caption{Chemical freeze-out parameters used in the manuscript for thermal model calculations.}
    \label{tab:chemicalparameters}
\end{table}

\begin{table}[h]
    \centering
    \begin{tabular}{c c c c}
        \toprule
        $\sqrt{s_{\rm{NN}}}$ (GeV) & $T_{\rm kin}$ (MeV) & $\langle \beta \rangle$ ($c$) & $n$ \\
        \midrule
        3    & $63\pm10$  &  ~$0.45\pm0.04$~  & 1 \\
        4.8  & $127\pm15$  & $0.39\pm0.05$  & 1 \\
        7.7  & $117\pm11$  & $0.45\pm0.05$  & ~~$0.5\pm0.3$~~ \\
        11.5 & $119\pm12$  & $0.46\pm0.05$  & $0.6\pm0.3$ \\
        19.6 & $114\pm12$  & $0.46\pm0.03$  & $0.9\pm0.2$ \\
        27   & $117\pm11$  & $0.48\pm0.03$  & $0.7\pm0.2$ \\        
        \bottomrule
    \end{tabular}
    \caption{Kinetic freeze-out parameters used in the manuscript for blast-wave predictions.}
    \label{tab:kineticparameters}
\end{table}
\section{Branching Ratio of ${}^{3}_{\Lambda}\rm{H} \rightarrow {}^{3}_{}\rm{He} + \pi^{-}$}

Although there are numerous model calculations on the branching ratio of ${}^{3}_{\Lambda}\rm{H} \rightarrow {}^{3}_{}\rm{He} + \pi^{-}$~\cite{Kamada:1997rv}, there is so far no direct experimental measurement. On the other hand, the relative branching ratio $R_3$, typically defined as

\begin{align}
    R_3 = \frac{B.R.({}^{3}_{\Lambda}{\rm{H}} \rightarrow {}^{3}_{}{\rm{He}} + \pi^{-})}{B.R.({}^{3}_{\Lambda}{\rm{H}} \rightarrow {}^{3}_{}{\rm{He}} + \pi^{-}) + B.R.({}^{3}_{\Lambda}{\rm{H}} \rightarrow d + p + \pi^{-})},
\end{align}

\noindent has been measured to good precision: $0.357 \pm ^{0.028}_{0.027}$~\cite{Eckert:2022dyz}. In order to make an estimate of the branching ratio and its associated uncertainty, we use a hybrid approach inspired by Ref.~\cite{Gazda:2023fow}, utilizing experimental measurements and theoretical constraints as follows:

 The decay modes of the \hyt include ${}^{3}{\rm{He}} + \pi^{-}$, $t + \pi^{0}$, $d + p + \pi^{-}$, $d + n + \pi^{0}$, 4-body mesonic decays, and non-mesonic decays. From theoretical calculations, it is generally agreed that the contributions from the 2-body and 3-body mesonic decays are dominant~\cite{Gazda:2023fow,Kamada:1997rv}. The 4-body mesonic decay branching ratio is estimated to be $0.9\%$ by Kamada et al.~\cite{Kamada:1997rv}, which we will take as our best estimate. For the non-mesonic decay branching ratio, Kamada et al. predicts a value of $1.7\%$. Golak et al.~\cite{Golak:1996hj} predicts a decay rate of $1.5\%$ of the free $\Lambda$, which also translates to a branching ratio of $1.7\%$ if we take the \hyt lifetime to be $237$ps~\cite{Eckert:2022dyz}. Thus we will take $1.7\%$ as our best estimate. As for the two-body and three-body mesonic decays, an assumption that is often taken is the isospin rule, which assumes that the branching ratio to $\pi^- + X$ is two times that of $\pi^0 + X$~\cite{Gazda:2023fow}. For the $\Lambda$ decay, $B.R.(\pi^- + X)/B.R.(\pi^0 + X) = 1.78$, hence we take the relative deviation from $2$, $10\%$ as an uncertainty in the application of this rule to the \hyt.

Taking into account all the information above, we calculate the branching ratio of ${}^{3}_{\Lambda}\rm{H} \rightarrow {}^{3}_{}\rm{He} + \pi^{-}$ using the formula:
\begin{align}
    B.R.({}^{3}_{\Lambda}{\rm{H}} & 
 \rightarrow {}^{3}_{}{\rm{He}} + \pi^{-}) \notag \\
    = & R_3 \times [1- B.R.({}^{3}_{\Lambda}{\rm{H}} \rightarrow {\rm{4\textrm{-} body\, mesonic)}} \notag \\
    & - B.R.({}^{3}_{\Lambda}{\rm{H}} \rightarrow {\rm{non\textrm{-} mesonic)]}} \times \frac{2}{3},
\end{align}

Accounting for the experimental uncertainty in $R_3$ and the uncertainty in the accuracy of the isospin rule, we arrive at our estimate:
\begin{align}
    B.R.({}^{3}_{\Lambda}{\rm{H}} \rightarrow {}^{3}_{}{\rm{He}} + \pi^{-}) = (23\pm3)\%.
\end{align}

\section{Effect of unstable nuclei feed-down on $(^{3}_{\Lambda}{\rm H}/\Lambda)/(t/p)$}

\begin{figure}
    \centering
    \includegraphics[width=0.9\linewidth]{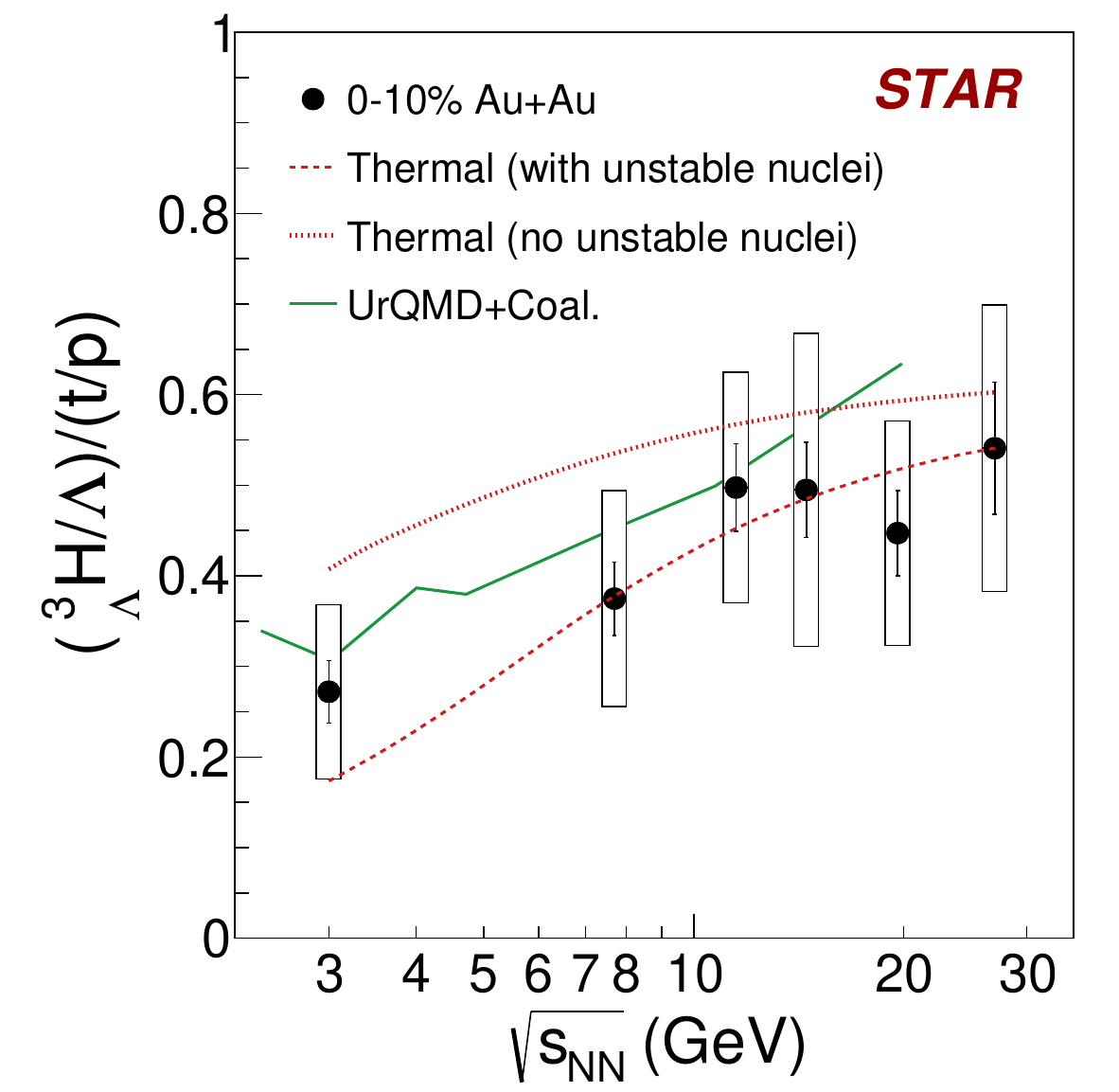}
    \caption{Collision energy dependence of the double ratio $(^{3}_{\Lambda}{\rm H}/\Lambda)/(t/p)$. The dashed and dotted red curves are from thermal model calculations considering and without considering unstable nuclei respectively~\cite{Vovchenko:2015idt}, while the green curve is based on the UrQMD model with  coalescence of hyperons and nucleons as an afterburner~\cite{Reichert:2022mek}. Boxes around data points indicate systematic uncertainties, and the vertical bars represent statistical uncertainties. The 13\% uncertainty in the ${}^{3}_{\Lambda}\rm{H}$ branching ratio is not shown. }
    \label{fig:nounstablenuclei}
\end{figure}

Figure~\ref{fig:nounstablenuclei} shows the collision-energy dependence of the double ratio $(^{3}_{\Lambda}{\rm H}/\Lambda)/(t/p)$. The data are compared with two configurations of the Thermal-FIST calculations: one including feed-down from unstable nuclei and one without such contributions.

Feed-down from unstable nuclei does not affect the ${}^{3}_{\Lambda}\rm{H}$ yield in the thermal model, because no known hypernuclei decay into ${}^{3}_{\Lambda}\rm{H}$. However, it does affect the triton yield through decays of excited nuclear states, for example ${}^{4}\mathrm{He}^{*} \rightarrow p + t$~\cite{Vovchenko:2020dmv}. As a result, the predicted triton yield increases when unstable nuclei are included, which leads to a reduction of the double ratio $({}^{3}_{\Lambda}{\rm H}/\Lambda)/(t/p)$ in the thermal-model calculation. Within the current experimental uncertainties, both configurations provide a reasonable description of the data.

\normalem
\bibliographystyle{apsrev4-2}
\bibliography{ref_supp}